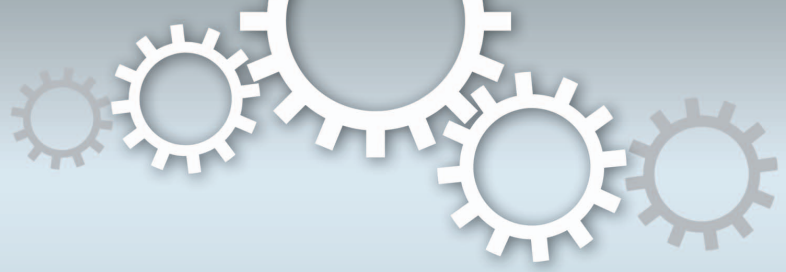

# Loops and autonomy promote evolvability of ecosystem networks

Jianxi Luo

Singapore University of Technology and Design, 20 Dover Drive, Singapore 138682.

The structure of ecological networks, in particular food webs, determines their ability to evolve further, i.e. evolvability. The knowledge about how food web evolvability is determined by the structures of diverse ecological networks can guide human interventions purposefully to either promote or limit evolvability of ecosystems. However, the focus of prior food web studies was on stability and robustness; little is known regarding the impact of ecological network structures on their evolvability. To correlate ecosystem structure and evolvability, we adopt the *NK* model originally from evolutionary biology to generate and assess the ruggedness of fitness landscapes of a wide spectrum of model food webs with gradual variation in the amount of feeding loops and link density. The variation in network structures is controlled by linkage rewiring. Our results show that more feeding loops and lower trophic link density, i.e. higher autonomy of species, of food webs increase the potential for the ecosystem to generate heritable variations with improved fitness. Our findings allow the prediction of the evolvability of actual food webs according to their network structures, and provide guidance to enhancing or controlling the evolvability of specific ecosystems.

The studies of ecological networks, in particular food webs, have focused on the structure[1–3] and dynamics of such networks[4,5], and their correlations[4–8]. Evolvability is the capacity to allow random but heritable variations of species which produce improvements from the status quo[9–11], thus it affects ecosystem dynamics. Such variations can be ignited by the changes of a single or a set of species that also require or cause co-adaptation of its direct and indirect prey and predators[6,12]. Therefore, a food web's evolvability at a certain point of time is conditioned by its feeding network structure at that time. In this paper, we investigate how specifically the variation in food web structure influences its evolvability. We aim to contribute to the literature on ecological network structure and dynamics, which have primarily focused on stability[4,5,13–16] but ignored evolvability, in the past.

Recently, several models, such as the cascade model[17], niche model[18], nested-hierarchy model[19], and the generalized cascade niche model[20], have been developed to generate networks that capture the key structural properties of empirical food webs using two empirical inputs, the number of species and connectance[21]. These models have also been used to investigate complex properties of food webs, such as stability[13–16,23–25] and robustness to species loss[26,27], as a result of food web structures. The evolvability of an ecosystem, i.e., its capacity to generate fitness-improving heritable variations[9,10], is complementary to stability and robustness which address the ability of the ecosystem to resist collapse or change with fitness loss. However, ecosystem evolvability has not been studied.

The food web model developed in this study incorporates the key mechanisms of food webs, including cascade hierarchy[17], niche[18], multidimensionality and intervality of niches[20,25], as well as rewiring[28,29] (i.e. predators switching to prey not previously consumed), which have been considered in previous models. On that basis, our model incorporates two *tuning* parameters to generate networks with gradual structural variations, to be used to investigate their evolvability given their varied structures.

The first tuning parameter is "predation diversity ($D$)", which denotes the degree to which a species' diet is non-specific to a downstream niche on the food chain (Fig. 1), implying the scope of diverse prey types or adaptive foraging choices[15,19]. For example, omnivores are fed on more than one trophic level, so tend to have high predation diversity[30] and possibly lead to feeding loops[13]. The diversity of prey types consumed was qualitatively examined in prior work[31], but has not yet been quantified. The second tuning parameter is "link density ($K$)", i.e. the average number of prey (or trophic links) per species in a food web. $K$ is an indicator of food web interaction complexity[5,14,15,20], and monotonically correlated with the *connectance* measure in other food web studies[1,21,26]. A lower link density indicates higher autonomy of species. $D$ and $K$ are adjusted to rewire networks to represent a variety of food webs with gradually varied structures, for the purpose to correlate structure variation with ecosystem evolvability. By contrast, the purpose of previous models was to accurately replicate actual food webs.

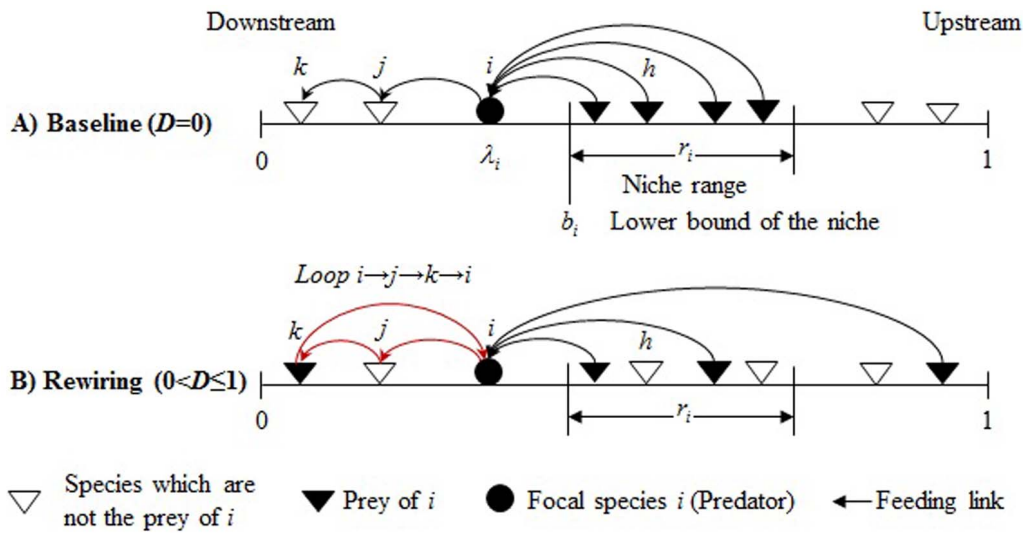

Figure 1 | **The rewiring food web model.** The red links form a loop.

The model has a baseline step (Fig. 1A) and a rewiring step (Fig. 1B) (mathematical details are provided in the Supplementary Information). The baseline step is modified from the previously introduced cascade niche models[18,20]. In the baseline scenario, before rewiring, predation diversity ($D$) is zero. Each of the $N$ species is assigned to a uniformly-distributed random position $\lambda_i$ along an axis ranging from 0 to 1. All prey of a species are in a continuous diet niche randomly located anywhere fully upstream to itself. The size of species $i$'s diet niche $r_i$ is associated with its total downstream range $1 - \lambda_i$ and link density $K$ (mathematical association is given in Supplementary Information). In contrast to the original niche model[18] in which a diet niche of a predator may overlap itself to allow cannibalism and loops, our setting can control and adjust the amount of loops to be generated by the later rewiring procedure (see Fig. 2 in the Results section). The baseline model generates networks that fully obey hierarchy and niche rules and does not allow feeding loops as all feeding links are oriented toward downstream.

When "predation diversity" is tuned greater than zero ($0 < D \leq 1$), a portion ($D$) of predating links of a species, which presumably go in its hypothesized niche, become non-specific to its predefined niche and may be rewired to species anywhere (Fig. 1B). A higher $D$ means a higher percentage of the predator's prey is outside the hypothesized single niche given in the baseline scenario. At the extreme $D = 1$, all species can consume prey anywhere on the axis. As a result, via rewiring, loops among species can emerge to the degree related to $D$. Fig. 1B demonstrates a loop to be formed between species $i$, $j$ and $k$ when a feeding link of species $i$ is rewired from $h$ to $k$. The rewired networks may have discontinuous diet ranges that also appear in the representations of actual food webs, as the result of 1) consuming multiple diet niches in one dimension of traits as the consumer's adaptive foraging or exploratory behaviors[15], or 2) mapping continuous diet niches in multiple dimensions of prey traits (such as body mass and movement speed)[20,22] to a one-dimensional axis (see Ref. 20 for detail explanations).

Recent studies of food web stability in response to simulated species loss have incorporated "rewiring"[28,29] to consider dynamics. In those models, following the removal of a species, some of its prey links may be rewired to new predators. Our use of rewiring is for a different purpose, which is to create food web-like networks with varied structures, similar to the rewiring method used to simulate "small-world" networks[32]. Herein, we are particularly interested in controlling and adjusting the amount of loops to be generated by the extent of rewiring.

In characterizing the structure variation in food webs, our primary lens is "feeding loops"[13]. We measure the general degree to which species in a food web predate or feed each other in loops as the percentage of trophic links that are included in a loop,

$$L = \sum_{i=1}^{M} e_i \Big/ M,$$

where $M$ is the number of links and $e_i = 1$ if link $i$ is in a loop and 0 otherwise. We term $L$ "degree of loops" or "loop degree". When $L = 0$, there is no loop in the network; the food web is acyclic. When $L = 1$, any trophic link is in at least one loop; the food web is fully cyclic. When $0 < L < 1$, the food web is partially cyclic; this is the most likely scenario in actual food webs.

The evolvability of model food webs with varied network structures is assessed using the $NK$ model framework[33–35], originally developed to study genome evolution. In a potential $NK$ framework for ecosystems (more model details are provided in Supplementary

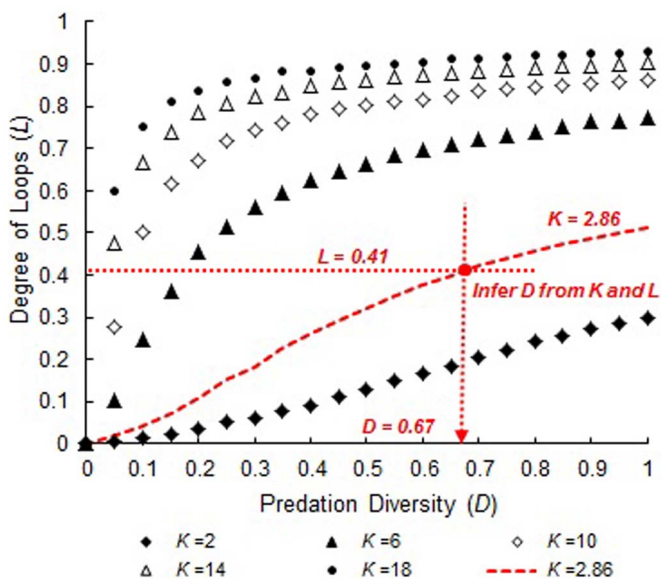

Figure 2 | **The impact of predation diversity and link density on loop degree.** For each given combination of inputs ($K$, $D$) with $N$ fixed at 100, we simulate an ensemble of 2,000 networks and calculate the average loop degree of the ensemble. The red dashed lines demonstrate a process to infer $D$ from empirically known $K = 2.86$ and $L = 0.41$.

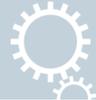

Information), a food web has $N$ species, each of which has $\omega_i$ possible mutation states and $K_i$ prey, i.e. predation links. In actual food webs, the biological states and changes of states of species are normally not discrete but continuous. For computational ease, hereafter we consider the simplest case that $\omega_i = 2$ for all $i$, and the state change of a species between 0 and 1 represents a general change due to either the species' endogenous mutation or adaption to environmental changes (such as climate or habit condition changes)[36]. Thus, the configuration of an ecosystem of $N$ species can be described by an $N$-digit string of 1 s or 0 s, denoted as $s_i = d_1 d_2 \cdots d_k \cdots d_N$, with $d_k = 0$ or 1, for $k = 1,2,3,\cdots,N$ and $i = 1,2,3,\cdots,2^N$. The combinatory space of $N$ species has total $2^N$ possible configurations. Different configurations have different degrees of fitness.

In the NK simulation, the fitness of a species is randomly drawn from uniform distribution [0,1], each time itself, its prey or predators change states. This setting represents that both predator and prey in a trophic link can mutually affect or depend on each other. The fitness of each configuration of the ecosystem is evaluated as the average of all species' fitness values. The fitness values of all $2^N$ configurations form the "fitness landscape" of an ecosystem. As the fitness of one species depends on the states of other species in a way shaped by the pattern structure of their feeding relationships, simulations create a mapping from the network structure of an ecosystem to its fitness landscape (for detailed procedure of landscape generation, please refer to Supplementary Information). Prior NK simulation analysis of general networks has shown that the varied topological patterns of interdependences (i.e. network links) can affect the shape characteristics of the fitness landscape[37], when the amount of interdependences is controlled.

On an NK landscape, if the fitness of a configuration (e.g., 001 when $N = 3$) is higher than that of any of its 1-mutant neighbor (e.g., 101,011,000), this configuration is considered a local peak. If a local peak's fitness is the highest among all configurations, it is also a global peak. The number of peaks of a fitness landscape indicates its "ruggedness", which is the opposite of "smoothness"[33–35]. On a highly rugged landscape where there are many local peaks, an average starting point will have a high chance to include a local peak in its neighborhood or be close to a local peak. Evolution via the mutation of a single species at a time will quickly reach and stabilize at a local peak closest to an average starting point, exhausting the potential of further fitness-increasing variation and evolution. In contrast, in a not-so-rugged or even single-peak landscape, it is more likely for the neighborhood of an average starting point to include a configuration with higher fitness due to the lack of local peaks. This indicates a higher chance for the ecosystem to generate and adapt to fitness-improving configuration variations.

Therefore, fitness landscape ruggedness of a food web implies a system constraint on its capacity to subsequently generate and adapt to configuration variations with higher fitness, i.e. evolvability. Therefore, we use the average number of peaks ($P$) on the fitness landscapes of a food web as a *reverse* indicator of its evolvability. Furthermore, as indicated in the NK model and simulation framework, the fitness landscape is only determined by the structure of feeding relationships, which is intrinsic to the ecological network. Note that, ecosystems with lower intrinsic evolvability determined by their network structures may also experience dramatic evolution. For instance, when extrinsic factors drive multiple or many species to mutate at the same time, the neighborhood of an average starting point will be wider, implying a higher chance to include varied configurations with improved fitness.

## Results

Tuning the parameters, including number of species ($N$), link density ($K$) and predation diversity ($D$), the model yields networks with gradually varied degrees of loops ($L$). A few regularities of the model-generated networks are identified. First, $L$ appears almost unaffected by changes in $N$ when $N$ is sufficiently large. That implies $L$ is a function of $K$ and $D$, i.e., $L = f(K, D)$. Specifically, $L$ increases with either $D$ or $K$ (Fig. 2) when $N$ is fixed. Furthermore, as the relationship between $K$ or $D$ and $L$ is monotonic, one can infer the nominal average predation diversity of an actual food web from the empirically measurable link density and loop degree. That is, $D = f^{-1}(L, K)$.

Given the pattern in Fig. 2, one can control the amount of loops to be generated in food webs by adjusting predation diversity $D$ and link density $K$. On that basis, we use the rewiring model to generate a wide spectrum of food webs with varied amount of loops and link density, and then use the NK model to simulate their fitness landscapes, and assess their evolvability. Our model settings ensure that network structure is the only determinant of the shape characteristics of the fitness landscape. The effects of all other factors are randomized.

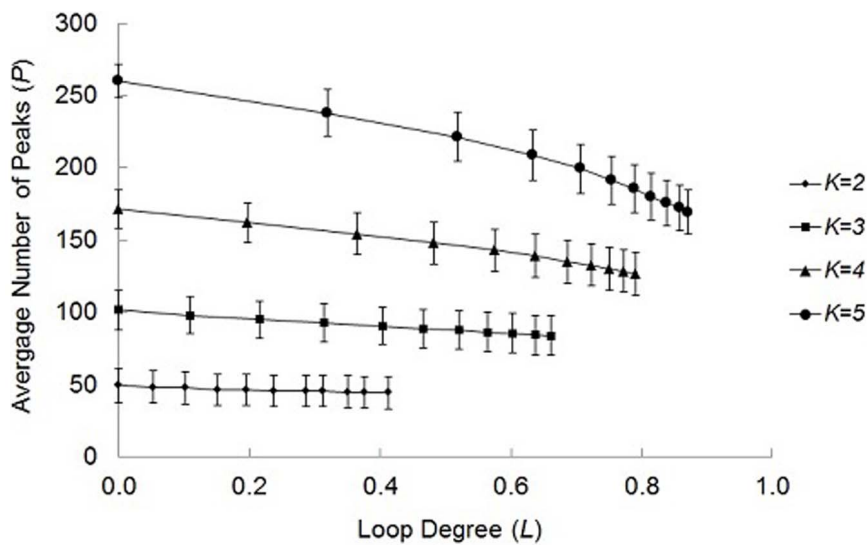

**Figure 3 | Loops and fitness landscape ruggedness, i.e. reverse indicator of evolvability.** In the simulations, for each given combination of inputs ($K, S$) and fixed $N = 12$, we simulate an ensemble of 2,000 networks and calculate their average loop degree. For each network in one ensemble, we generate 200 NK fitness landscapes and calculate the average number of peaks of theirs. Error bars represent standard deviations.

Our results show that an increase in the degree of loops ($L$) of a food web reduces the average number of peaks ($P$), i.e., landscape ruggedness, when holding fixed link density ($K$) (Fig. 3). This indicates feeding loops promote evolvability. Conversely, less cyclic food webs will be less evolable. Furthermore, due to the positive monotonic relationship between predation diversity and loop degree (Fig. 2), one can further infer that predation diversity promotes evolvability.

Two detailed patterns on the loop-ruggedness relationship are noteworthy. First, the decrease in $P$ is fairly linear with respect to $L$. The Pearson correlation coefficients for $P$ and $L$ range from 0.980 for $K = 5$ to 0.983 for $K = 2$. Second, when $K$ is higher, $P$ decreases faster with the increase in $L$. The slope of the $P$-$L$ linear regression curve is $-105.7$ for $K = 5$ and much higher than the slope of $-11.6$ for $K = 2$. This indicates a reinforcing effect of link density on the promoting effect of feeding loops on food web evolvability.

Our results further show that fitness landscape ruggedness increases with increases of $K$, when $L$ is fixed. This indicates the link density of a food web, i.e. interaction complexity, limits its evolvability. The positive density-ruggedness relationship is also reinforced by the decrease in loop degree or predation diversity, shown by the expanding gaps between the lines for various $K$ values as $L$ decreases in Fig. 3. Because lower link density implies higher autonomy of species, we can conclude that feeding autonomy promotes food web evolvability.

Note that the promoting effect of loops ($L$) and the limiting effect of link density ($K$) are coupled to co-determine evolvability. For instance, an ecosystem with higher $L$ can be less evolvable than another ecosystem with a lower $L$, if it has a much higher $K$ to drive down the overall evolvalibity. One example is that, the right most point on curve $K = 5$ in Fig. 3, despite having a much higher $L$, has a lower evolvability (i.e. higher $P$) than the left most point on curve $K = 3$.

## Discussion

To summarize, loop degree and link density co-determine evolvability. Specifically, the degree to which species feeds (or predate) each other in loops, and the autonomy of species in the trophic network, promote the evolvability of ecosystems. Conversely, hierarchical or acyclic feeding (or predation) relationships and interaction complexity limit ecosystem evolvability. With this understanding (and the specified influences in Fig. 3), one can predict and compare the evolvability of actual food webs, once their network structures are known. The network structures, in terms of link density and loop degree, are empirically measurable. They can also be used to infer predation diversity (as demonstrated in Fig. 2).

A better understanding of the impact of food web structure on its evolvability can guide the strategies of human interventions for either higher or lower evolvability. Furthermore, the density-evolvability and loop-evolvability relationships are generalizable for broadly defined ecosystems other than food webs, such as technological and economic systems[38] that also have many interdependent co-evolving elements, to predict their evolvability or to (re-)design their system architectures for the purpose of controlling or adjusting evolvability.

### Acknowledgments

The author thanks Eric Lucas for *NK*-related computer programming and simulations.

### Additional information

**Supplementary information** accompanies this paper at http://www.nature.com/scientificreports

**Competing financial interests:** The authors declare no competing financial interests.